\documentclass[12pt]{article}
\pdfoutput=1

\usepackage[utf8]{inputenc}
\usepackage[margin=1in]{geometry}
\usepackage[tbtags]{amsmath} 				
\usepackage{amssymb, mathrsfs}			
\usepackage{physics}
\usepackage{graphicx}			
\usepackage[makeroom]{cancel}	
\usepackage{bm}								
\usepackage[
      colorlinks=true,
      linkcolor=blue,
      urlcolor=blue,
      filecolor=black,
      citecolor=red,
      pdfstartview=FitV,
      pdftitle={},
      pdfauthor={Michael Gutperle,Christoph Uhlemann},
      bookmarksopen=true
      ]{hyperref}

\usepackage{cite}

\usepackage{graphicx,calc}
\newlength\myheight
\newlength\mydepth
\settototalheight\myheight{Xygp}
\def\cN{{\cal N}}

\def\Re{{\rm Re \,}}
\def\Im{{\rm Im \,}}

\def\det{{\rm det \,}}

\def\ZZ{{\mathds{Z}}}

\renewcommand{\Im}{\operatorname{Im}}
\renewcommand{\Re}{\operatorname{Re}}

\DeclareMathOperator{\Vol}{Vol}
\DeclareMathOperator{\vol}{vol}

\newcommand{\RAdS}{\ensuremath{R_\mathrm{AdS}}}
\newcommand{\RS}{\ensuremath{R_{\mathrm{S}^5}}}

\newcommand{\GammaAdS}{\ensuremath{\Gamma_{\mathrm{AdS}}}}

\newcommand{\GammaChi}{\ensuremath{\Gamma_{\vec{\chi}}}}

\usepackage{sectsty}
\sectionfont{\large}

\renewcommand{\thefootnote}{\fnsymbol{footnote}}
\renewcommand{\thanks}[1]{\footnote{#1}}
\newcommand{\starttext}{
\setcounter{footnote}{0}
\renewcommand{\thefootnote}{\arabic{footnote}}}
\renewcommand\({\begin{equation}}		
\renewcommand\){\end{equation}}
\renewcommand{\epsilon}{\varepsilon}	

\usepackage{dsfont}

\newcommand{\RR}{\mathds{R}}

\numberwithin{equation}{section} 		

\numberwithin{equation}{section}

\renewcommand{\Im}{\operatorname{Im}}
\renewcommand{\Re}{\operatorname{Re}}

\long\def\symbolfootnote[#1]#2{\begingroup%
\def\thefootnote{\fnsymbol{footnote}}\footnote[#1]{#2}\endgroup}

\begin{document}
\setlength{\baselineskip}{16pt}

\starttext
\setcounter{footnote}{0}

\begin{flushright}
\today\\
LCTP-20-07
\end{flushright}

\bigskip

\begin{center}

{\Large \bf  Janus on the Brane}

\vskip 0.4in

{\large  Michael Gutperle${}^{ \includegraphics[height=.2in]{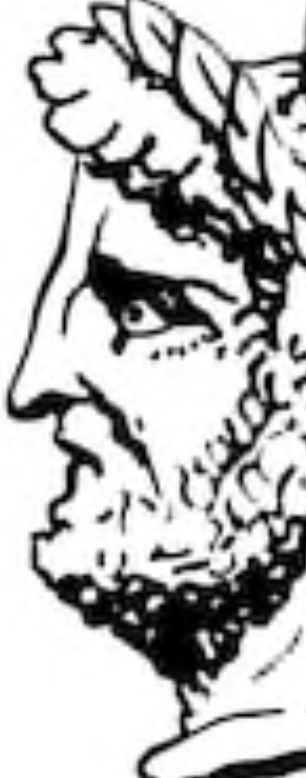}}$ and Christoph F. Uhlemann${}^{ \includegraphics[height=.2in]{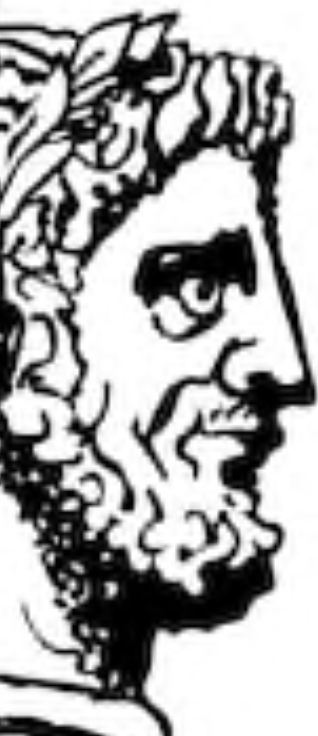}}$ }

\vskip 0.2in

${}^{ \includegraphics[height=.2in]{janus-r2}}${\sl Mani L. Bhaumik Institute for Theoretical Physics}\\
{\sl Department of Physics and Astronomy }\\
{\sl University of California, Los Angeles, CA 90095, USA} 

\medskip

${}^{ \includegraphics[height=.2in]{janus-l2}}${\sl Leinweber Center for Theoretical Physics, Department of Physics}\\
{\sl University of Michigan, 450 Church Street, Ann Arbor, MI 48109-1020, USA}

\bigskip

\end{center}
 
\begin{abstract}
\setlength{\baselineskip}{16pt}

We present a non-supersymmetric deformation of probe branes describing conformal defects of codimension two in AdS/CFT. 
The worldvolume of the probe branes is deformed from $AdS_{p}\times S^1$ embedded in an $AdS_{p+2} \times \mathcal M^{D-p-2}$ background to an embedding of Janus form, which uses an $AdS_{p-1}$ slicing of $AdS_p$ and in which the brane bends along the slicing coordinate.
In field theory terms this realizes conformal interfaces on codimension-two defects.
We discuss these ``Janus on the brane'' solutions for $AdS_3\times S^1$ D3-branes in the  $AdS_5\times S^5$ solution of Type IIB, realizing interfaces on surface defects in $\mathcal N=4$ SYM, and show that similar solutions exist for probe branes in $AdS_{p+2}\times S^{9-p}$ vacua of M-theory and in the $AdS_6\times S^4$ solution of massive Type~IIA.
\end{abstract}

\setcounter{equation}{0}
\setcounter{footnote}{0}

\newpage

\section{Introduction}

The study of defects and interfaces  in field theories is of considerable interest.  One of the first examples in holography was the Janus solution \cite{Bak:2003jk} which is a deformation of the $AdS_5\times S^5$ solution of Type IIB supergravity describing a planar  interface in $\cN=4$  SYM across which the coupling constant jumps. Subsequently, Janus solutions have been  generalized in many ways, see e.g. \cite{Clark:2005te,DHoker:2007zhm,DHoker:2009lky,Chiodaroli:2009yw,Bobev:2013yra,Pilch:2015dwa,Gutperle:2017nwo,Assel:2018vtq,Guarino:2019oct,Guarino:2020gfe,Bobev:2020fon}. Another way to construct defects in holographic theories is by embedding   probe branes in the supergravity dual and neglecting their backreaction \cite{Karch:2000gx}.  This approximation is often justified if the number of probe branes is  small compared to the number of branes which created the background  spacetime.   
Defects of various dimensions and in various field theories are described holographically by probe branes with $AdS_{p}\times S^q$  worldvolume, where the $AdS_{p}$ is embedded inside the $AdS$ part of the background and the $S^q$ can either be embedded inside the $AdS$ part or in the internal space \cite{DeWolfe:2001pq,Bachas:2001vj,Constable:2002xt,Aharony:2003qf,Fujita:2011fp}. 
In many cases BPS defects can be realized, which preserve part of the background supersymmetries and may involve additional worldvolume fluxes.

In this paper we present a deformation of defect probe branes  inspired by the Janus-within-Janus solution of \cite{Hirano:2006as}. The original Janus solution  \cite{Bak:2003jk} is  based on an $AdS_4$ slicing of $AdS_5$, with the dilaton depending on the slicing coordinate.  In  \cite{Hirano:2006as}  this solution was generalized by using an $AdS_3$ slicing for $AdS_4$ and making the dilaton dependent also on the second slicing coordinate.  Consequently, this solution describes a defect within a defect. We apply this idea to probe branes with an $AdS_{p}$ worldvolume by using an $AdS_{p-1}$ slicing of $AdS_{p}$ and making the embedding of the branes into the background spacetime  dependent on the slicing coordinate. Hence we name the resulting solutions  ``Janus on the brane''. These embeddings describe codimension-two defects in the dual field theories with an interface on the defect across which certain parameters characterizing the defect jump. The solutions generically break all supersymmetries preserved by the undeformed defect.
 
The remainder of this paper  is organized as follows: In section \ref{sec2} we discuss, as  our main example, two-dimensional surface defects in $\cN=4$ SYM which are described by probe D3-branes with $AdS_3\times S^1$ worldvolume in $AdS_5\times S^5$. We find numerical and perturbative Janus-on-the-brane solutions and discuss their interpretation as surface defects in $\cN=4$ SYM.  In section \ref{sec3} we generalize the construction to probe M2-branes with $AdS_2\times S^1$ worldvolume in the $AdS_4\times S^7$ vacuum of M-theory, probe M5-branes with $AdS_5\times S^1$ worldvolume in $AdS_7\times S^4$, and probe D4-branes with $AdS_4\times S^1$ worldvolume in the $AdS_6\times\hat S_4$ Brandhuber-Oz solution of massive Type IIA. We close with a discussion and outlook in section~\ref{sec4}.

\section{Surface operators in \texorpdfstring{$\cN=4$}{N=4} SYM}\label{sec2}

Surface operators of disorder type in $\cN=4$ SYM were constructed in \cite{Gukov:2006jk} in a semiclassical approximation.  The half-BPS surface operators preserve a $PSU(1,1|2)\times PSU(1,1|2)$ subgroup of the $PSU(2,2|4)$ superconformal symmetry of $\cN=4$ SYM.
In \cite{Gukov:2006jk} the surface operator is realized as  a vortex configuration and a singular gauge field transverse to $\Sigma_2= R^{1,1}$ in $R^{1,3}$.  In this paper we will use an equivalent description introduced in \cite{Drukker:2008wr}, which maps  four-dimensional Minkowski space conformally to $AdS_3 \times S^1$ with metric
\begin{equation}\label{ads3s1}
ds^2= ds^2_{AdS_3}+ d\psi^2
\end{equation}
The surface operator is located at the conformal boundary of $AdS_3$ and corresponds to a non-trivial state on $AdS_3\times S^1$.
It is characterized by a choice of Levi-Group $L= \prod_{n=1}^m U(N_n) \in U(N)$ and a vortex configuration for the gauge field
\begin{equation}
A = 
\left(
\begin{array}{cccc}
  \alpha_1 {\bf 1}_{N_1}&  0  & \cdots &  0 \\
0   &  \alpha_2 {\bf 1}_{N_2}  & \ddots  & \vdots\\
   \vdots &  \ddots  &  \ddots & 0 \\
0  &\cdots& 0 & \alpha_m{\bf 1}_{N_m}  \\
\end{array}
\right) d\psi
\end{equation}
Among the six scalars of $\cN=4$ SYM only the combination  $\Phi ={1\over \sqrt{2}} (\phi^1+i \phi^2)$ is non-vanishing and has the following behavior
\begin{equation}\label{phifield}
\Phi = {e^{- i \psi} \over \sqrt{2}}
\left(
\begin{array}{cccc}
( \beta_1 + i \gamma_1)  {\bf 1}_{N_1} & 0   &\cdots   & 0\\
0   & ( \beta_2 + i \gamma_2) {\bf 1}_{N_2} &  \ddots & \vdots \\
  \vdots &  \ddots   & \ddots   & 0\\
  0&\cdots & 0& ( \beta_{m}  + i \gamma_m)  {\bf 1}_{N_m} 
\end{array}
\right)
\end{equation}
There is a further set of parameters $\eta_n,n=1,2,\cdots,m$ specifying theta angles for unbroken $U(1)$ factors.

In this paper  we will use the holographic description of  surface operators  as probe branes in $AdS_5\times S^5$ \cite{Drukker:2008wr,Gomis:2007fi}\footnote{Backreacted  Type IIB solutions were constructed in \cite{Gomis:2007fi}, based on a double analytic continuation of LLM solutions \cite{Lin:2004nb,Lin:2005nh}.}. 
With an $AdS_3\times S^1$ slicing  of $AdS_5$ and an $S^1\times S^3$ slicing of $S^5$ the $AdS_5\times S^5$ background metric reads
\begin{equation}\label{ads5met}
ds^2=   du^2 +\cosh^2 u \; ds_{AdS_3}^2 + \sinh^2 u\;  d\psi^2+ d\theta^2 + \cos^2\theta \; ds_{S^3}^2 + \sin^2\!\theta\, d\phi^2
\end{equation}
with $u\in [0,\infty)$.
The conformal boundary at $u\rightarrow\infty$ is $AdS_3\times S^1$, and this choice is natural for describing $\cN=4$ SYM on $AdS_3\times S^1$.
The four-form potential is given by\footnote{This $C_4$ differs from the choice in \cite{Drukker:2008wr} by a gauge transformation. It is regular at $u=0$
and leads to the correct anomaly, to be discussed briefly in sec.~\ref{sec:exp}. We thank Kristan Jensen for pointing this out to us.}
\begin{equation}
C_4 = ({\cosh^4\!u}-1) \;  \vol_{AdS_3} \wedge\, d\psi+\ldots
\end{equation}
where the dots denote components along the $S^5$ and $\vol_{AdS_3}$ is the volume form of unit-radius $AdS_3$.
As shown in \cite{Drukker:2008wr}, a  probe D3-brane  with worldvolume parameterized by the $AdS_3$ coordinates and $ \psi$ and the embedding 
\begin{align}\label{constemb}
u&=u_0 & \phi &=-\psi+ \psi_0 & \theta&= \frac{\pi}{2}
\end{align}
extremizes the action
\begin{equation}
S_{\rm D3}=  T_{\rm D3} \int d^4 \xi \sqrt{-det(\gamma_{ab} + F_{ab})} -T_{\rm D3} \int C_4 
\end{equation}
and preserves the same $PSU(1,1|2)\times PSU(1,1|2)$ superalgebra  as the surface operator.
A single D3-brane corresponds to a Levi group $U(1) \times U(N-1)$, and the relation to the parameters of the defect operator in $\cN=4$ SYM proposed in \cite{Drukker:2008wr} is
 \begin{align}\label{idena}
\beta+ i\gamma &= {\sqrt{\lambda}\over 2\pi} \sinh u_0 e^{i \phi_0}
&
\alpha&=\oint \frac{A}{2\pi}
&
\eta&=\oint \frac{\tilde A}{2\pi}
\end{align}
For any non-zero $u_0$ the parameter $\beta+i\gamma$ is of $\mathcal O(\sqrt{\lambda})$.
More general Levi groups $L$ can be realized by considering multiple D3-brane probes at different locations $u$. The backreaction can be neglected as long as the number of probe branes is small compared to~$N$.

\subsection{Janus on the D3-brane}\label{sec:janus}

For the Janus-on-the-brane configurations to be discussed in the following it is convenient to further foliate $AdS_3$ by $AdS_2$ slices, such that the $AdS_3$ metric in  (\ref{ads5met}) is given by
\begin{equation}
ds_{AdS_3}^2= d\xi^2 +\cosh^2\!\xi \,ds^2_{AdS_2}
\end{equation}
with $\xi\in\mathds{R}$.
The $AdS_2$ slices may be taken either as Poincar\'{e} or global (Euclidean) $AdS_2$.
The former case will describe an interface $\RR$ on a surface operator supported on the two copies of Poincar\'e $AdS_2$ obtained for $\xi\rightarrow\pm \infty$ and joined at their boundaries, which is conformally related to $\RR^2$.
The latter case will describe an $S^1$ interface on a surface operator supported on two copies of global $AdS_2$ joined at their boundaries, which is conformally related to $S^2$.
An ansatz for embeddings that preserve the $AdS_2$ isometries, corresponding to the defect conformal symmetry of the one-dimensional interface, is then obtained by allowing the $AdS_3\hookrightarrow AdS_5$ slicing coordinate $u$ to depend on the $AdS_2\hookrightarrow AdS_3$ slicing coordinate $\xi$
\begin{align}\label{eq:D3-embedding}
 u&=u(\xi) & \phi&= \phi(\psi) & \theta &= {\pi\over 2}
\end{align}
The action for a D3-brane embedded in such a way is given by
\begin{align}\label{sdthree}
S_{\rm D3} = T_{\rm D3}{\rm Vol}_{AdS_2}\int d\xi d\psi  \cosh^2\!\xi   \Bigg[&\cosh^2\!u\sqrt{ \big(\sinh^2\!u + \dot \phi^2 \big) \big( \cosh^2\!u+   (u')^2 \big)}
- \cosh^4\! u+1\Bigg]
\end{align} 
where ${\rm Vol}_{AdS_2}$ is the (renormalized) volume of $AdS_2$.
The Euler-Lagrange equation  for $\phi$  derived from this action is
\begin{eqnarray}
\sinh^2\! u(\xi)\,  {d^2 \phi(\psi) \over d\psi^2}&=&0 
\end{eqnarray}
It is solved by a generic linear function. The solutions we will use in the following are\footnote{Solutions with constant $u=u_0$ exist for $\phi=a\psi+\phi_0$ if either $a^2=1$ and $u_0$ arbitrary, or if $a^2\leq \tfrac{1}{9}$ and $\cosh^2 u_0=\frac{9}{8}(1-a^2)$. We focus here on deformations of the BPS embeddings with $a^2=1$.}
\begin{align}\label{bpswind}
 \phi &=\phi_0-\psi
\end{align}
Using them in the Euler-Lagrange equation for $u(\xi)$ leads to
\begin{eqnarray}\label{eqofm1}
 u''-5 u'^2\tanh u+2\tanh \xi\left(u'+\sech^2\! u \;  u'^3\right) - 2 \sinh (2u)&&\nonumber\\
+(4 \sinh u + 4 \; \sech u\, \tanh u \; u'^2) \sqrt{\cosh^2 u+ u'^2}  &=&0
\end{eqnarray}

 \begin{figure}[!t]
  \centering
  \includegraphics[width=80mm]{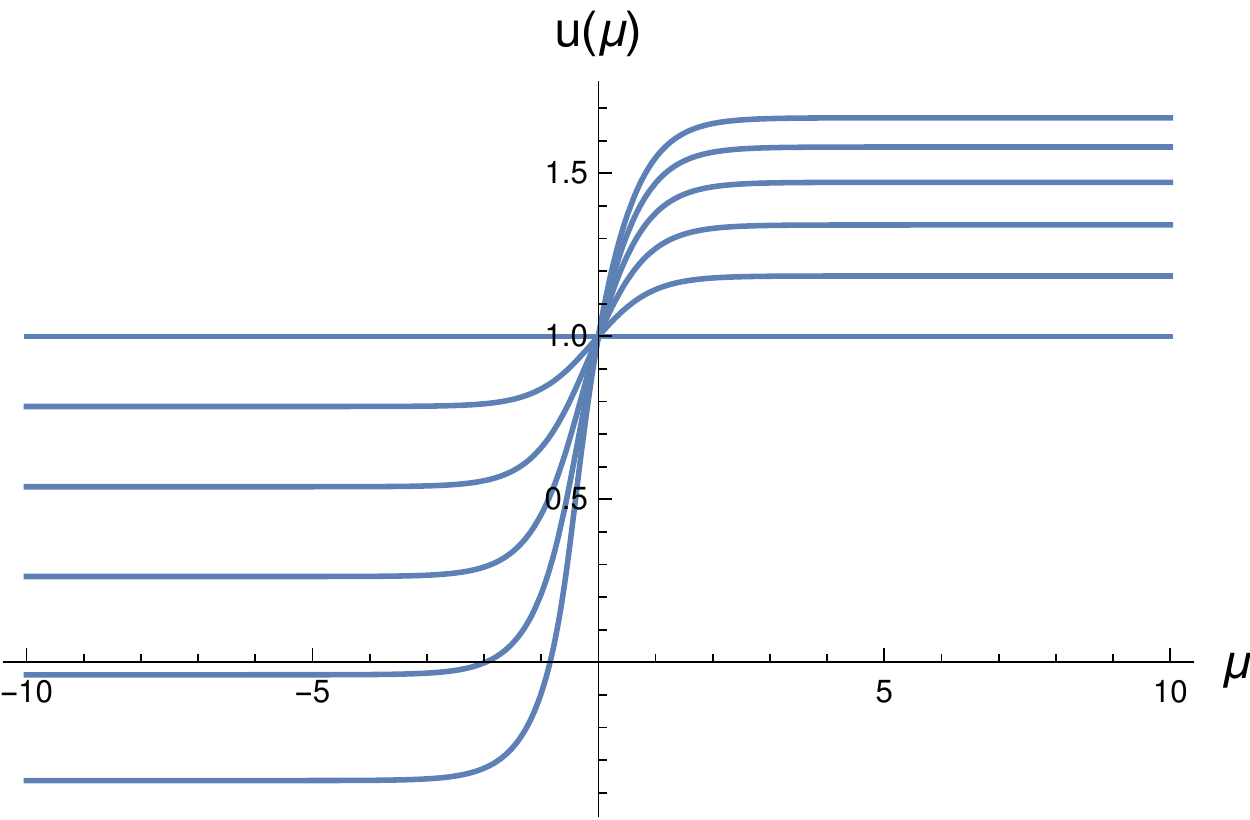}
  \caption{Sample of solutions to (\ref{eqofm1}) with $u(0)=1$ and $u'(0)\in \lbrace 0,0.2,0.4,\ldots, 1\rbrace$.
  Their qualitative behavior is well captured by the leading-order perturbative solution (\ref{lins}).}
  \label{fig:plot1}
\end{figure}

A sample of numerical solutions is shown in figure \ref{fig:plot1}.
For the non-constant solutions the embedding coordinate $u(\xi)$ approaches different values as the $AdS_2$ slicing  coordinate approaches $\xi\to \pm \infty$. This behavior is reminiscent of  Janus solutions, where it is now the embedding coordinate $u$ which jumps. 
Note that the coordinate $u$ in the $AdS_5\times S^5$ metric (\ref{ads5met}) only takes non-negative values -- 
the solutions where $u(\xi)$ changes sign can be interpreted as brane embeddings consisting of two branches with $u=|u(\xi)|$ and phase shifts in (\ref{bpswind}) differing by $\pi$ that are joined at the location where $u(\xi)$ changes sign.\footnote{%
The D3-brane wraps a curve with winding number $(1,1)$ in the torus $S^1_\psi\times S^1_\phi$. At $u=0$ the $S^1_\psi$ degenerates but $S^1_\phi$ does not, so the D3-brane does not cap off.
The $AdS_5\times S^5$ metric near $u=0$ is $ds^2\approx du^2+u^2d\psi^2+d\xi^2+d\phi^2+\ldots$. 
An embedding where $u(\xi)$ changes sign at $\xi=\xi_0$, with $u=|u(\xi)|$ and $\phi=\phi_0- \psi+\pi\Theta(-u(\xi))$, near $\xi_0$ describes a straight line through the origin of the plane $\RR^2_{(u,\psi)}$ for each $\phi$.}
The natural generalization of the identification between the parameters of the surface operator and those of the D3-brane embedding in (\ref{idena}) is
\begin{align}\label{beta-gamma}
 (\beta+i\gamma)_{\pm}&=\lim_{\xi\rightarrow \pm\infty}\frac{\sqrt{\lambda}}{2\pi}\sinh u(\xi)\, e^{i\phi_0}
\end{align}
The Janus-on-the-brane solution thus describes two 1+1 dimensional surface defects with different parameters $(\beta+i \gamma)_\pm$ glued together at a $0+1$ dimensional interface. As shown in appendix~\ref{sec:job-susy} these solutions break all supersymmetries.

We have not found an analytic solution to (\ref{eqofm1}), but a perturbative solution for small deviations from the supersymmetric embedding with constant $u$ can be found straightforwardly. In view of the identification with field theory parameters in (\ref{beta-gamma}), a natural ansatz for a perturbative expansion is
\begin{equation}\label{pertexp}
\sinh u(\xi) = \sinh u_0 + \epsilon  u_1(\xi) + {1\over 2} \epsilon^2 u_2(\xi) +{1\over 3!} \epsilon^3 u_3(\xi) +\ldots 
\end{equation}
Solving (\ref{eqofm1}) order by order in $\epsilon$ leads at leading order to the equation
\begin{equation}\label{probelin}
u_1''(\xi)+ 2 \tanh\xi \; u_1'(\xi)  =0
\end{equation}
which is solved by
\begin{equation}\label{lins}
u_1 = \alpha_1 \tanh \xi+\alpha_2
\end{equation}
Since $\alpha_1$ and $\alpha_2$ can be absorbed into a redefinition of $\epsilon$ and $u_0$, respectively, we set 
\begin{align}
 \alpha_1&=1 & \alpha_0&=0
\end{align}
in the following. By similar reasoning the integration constants appearing in the higher-order solutions can be fixed, by demanding that 
\begin{align}
\lim_{\xi\rightarrow\pm\infty}\sinh u(\xi)&=\sinh u_0\pm\epsilon 
\end{align}
That is, the higher orders should not redefine the expansion parameter (the difference between $(\beta+i\gamma)_\pm$) or $u_0$ (the average of $(\beta+i\gamma)_\pm$).
This leads to
\begin{align}\label{nonlin}
 u_2(\xi)&=0
 \nonumber\\
 u_3(\xi)&=\frac{4 \cosh^2\!\xi+3}{5\cosh^4\!\xi\,  \cosh^4\! u_0} \tanh\xi
 \nonumber\\
 u_4(\xi)&=2\sinh u_0\frac{19 - 2\cosh^2\! \xi -8 \cosh^4 \!\xi}{5\cosh^6\!\xi \cosh^6\!u_0}
\end{align}
and higher-order terms can be obtained straightforwardly.
The expansion for $u(\xi)$ is invariant under simultaneous sign reversal of $\epsilon$ and $\xi$, which dictates the parity of $u_i$.

\subsection{Defect expectation value}\label{sec:exp}

The holographic description of the surface operator $\mathcal O_\Sigma$ allows to compute many observables at strong coupling, such as correlation functions, entanglement entropy or the central charge associated with the conformal defect. In this section we focus on the expectation value, which is computed from the D3-brane on-shell action via
\begin{align}\label{eq:exp-O}
 \langle \mathcal O_\Sigma\rangle&= e^{-S_{\rm D3, on-shell}}
\end{align}
For a two-dimensional defect there can be a conformal anomaly, which, as discussed in \cite{Drukker:2008wr}, is characterized by three curvature invariants with independent coefficients. 
For a defect supported on $\RR^2$ the curvature invariants vanish, while for a defect on $S^2$ one of them is non-zero.
The anomaly was studied and shown to be non-vanishing in \cite{Jensen:2018rxu,Chalabi:2020iie} (see also \cite{Gentle:2015ruo}), 
amending the previous conclusion in \cite{Drukker:2008wr} that the anomaly vanishes.\footnote{
We thank the authors of \cite{Jensen:2018rxu,Chalabi:2020iie} for making us aware and explaining their work to us.}

We now compute the defect contribution to the expectation value.
The Janus-on-the-brane solutions asymptotically approach the constant embedding, and the integrand in (\ref{sdthree}) falls off sufficiently fast for large $|\xi|$ that there are no new divergences associated with large $|\xi|$; the conformal anomaly is unchanged.
For the perturbative solution  (\ref{pertexp}) one can calculate the on shell action as a power series in $\epsilon$.
The first terms following from (\ref{lins}) and (\ref{nonlin}) (and further terms in the expansion) are
\begin{align}\label{eq:SD3-onshell}
S_{\rm D3}=S_{\rm D3}^{(0)}+ 2 \pi T_{\rm D3}  & {\rm Vol}_{AdS_2}  \Bigg\{
\epsilon ^2
-\frac{2\epsilon ^4}{15\cosh^4\!u_0}
-\frac{4 \epsilon ^6 \left(75 \sinh^2\!u_0-23\right)}{1575\cosh^8\!u_0}
\nonumber\\&
-\frac{2 \epsilon ^8 \left(375375 \sinh^4\!u_0-564330 \sinh ^2\!u_0+55751\right)}{3378375\cosh^{12}\!u_0}
+\mathcal O(\epsilon^{10})
\Bigg\}
\end{align}
where $S_{\rm D3}^{(0)}$ is the action for the undeformed defect.
As expected it is invariant under $\epsilon\rightarrow -\epsilon$.
In general a finite on-shell action is obtained by including a hierarchy of holographic counterterms associated with the conformal boundaries of the $AdS$ slices of various dimensions (as discussed for example in \cite{Andrade:2011nh}).
Including counterterms on the boundary of the $AdS_2$ slices leads to the renormalized volumes of Poincar\'e and global $AdS_2$,
\begin{align}\label{eq:VolAdS2}
 {\rm Vol}_{AdS_2}&=-2\pi \quad (\text{global AdS$_2$}) & {\rm Vol}_{AdS_2}&=0 \quad (\text{Poincar\'e AdS$_2$})
\end{align}
The renormalized expectation value vanishes for an interface $\RR$ separating two copies of Poincar\'e $AdS_2$.
For an $S^1$ interface separating two copies of global $AdS_2$ it does not necessarily vanish.
The contribution from the undeformed defect, $S_{\rm D3}^{(0)}$ in (\ref{eq:SD3-onshell}), is divergent and encodes the anomaly discussed in \cite{Jensen:2018rxu,Chalabi:2020iie}. The holographic counterterms needed to render it finite break the bulk diffeomorphisms corresponding to conformal transformations on the boundary and introduce scheme dependence.
This is related to the availability of finite counterterms on the boundary of $AdS_3$:
cutting off the integral in (\ref{sdthree}) at large $|\xi|$, 
and following the logic for the holographic renormalization of probe branes of \cite{Karch:2005ms},
one may supplement the action by boundary terms of the form $\sqrt{\gamma}R_\gamma f(u)$, where $\gamma$ is the induced metric on the cut-off surface, $R_\gamma$ its Ricci scalar, and $f$ an arbitrary function of $u$.
A scheme can be fixed by demanding the on-shell action to vanish for arbitrary supersymmetric constant embeddings.
For an  $S^1$ interface separating two copies of global $AdS_2$ we then find
\begin{align}\label{eq:cO-J}
 \langle \mathcal O_\Sigma\rangle=\exp\Big[
 2N\Big\{&
\epsilon ^2
-\frac{2\epsilon ^4}{15\cosh^4\!u_0}
-\frac{4 \epsilon ^6 \left(75 \sinh^2\!u_0-23\right)}{1575\cosh^8\!u_0}
\nonumber\\&
-\frac{2 \epsilon ^8 \left(375375 \sinh^4\!u_0-564330 \sinh ^2\!u_0+55751\right)}{3378375\cosh^{12}\!u_0}
+\mathcal O(\epsilon^{10})\Big\}\Big]
\end{align}
We used the relation $T_{\rm D3}=N/(2\pi^2)$ for unit-radius $AdS_5$ \cite{Drukker:2008wr}, and $u_0$ is related to the field theory parameters characterizing the defect by (\ref{idena}).

\subsection{Janus Interface in Field Theory}

We will consider the simplest case of a Janus defect, which is a deformation of the scalar field defect (\ref{phifield}) with $\alpha_i$ and $\eta_i$ vanishing, from the field theory perspective.
For $\cN=4$ SYM with only a single complex field $\Phi={1\over \sqrt{2}} (\phi^1+ i \phi^2)$ nontrivial, the action reduces to \cite{Drukker:2008wr} 
\begin{equation}\label{eq:N4-action}
S= {1\over g_{\rm YM}^2} \int d^4x \sqrt{g}\,{\rm tr} \left( |D\Phi|^2+ {R\over 6} |\Phi|^2\right)
\end{equation}
The scalars are conformally coupled, which leads to the second term. For an $AdS_3\times S^1$ background (\ref{ads3s1}) we have $R=-6$ and the equation of motion becomes
\begin{equation}
D^2 \Phi + \Phi=0
\end{equation}
It is satisfied for the surface defect scalar field given in (\ref{phifield}). A  Janus like deformation of the surface defect in $AdS_3\times S^1$   can be obtained by using an $AdS_2$ slicing of $AdS_3$
\begin{equation}
ds^2 = d\xi^2 + \cosh^2\!\xi \,ds^2_{AdS_2} + d\psi^2
\end{equation}
and allowing the parameters $\beta_i, \gamma_i$ in the scalar field $\Phi$ defined in (\ref{phifield}) to depend on the slicing coordinate  $\xi$, leading to
\begin{equation}\label{phifield-J}
\Phi = {e^{- i \psi} \over \sqrt{2}}
\left(
\begin{array}{ccc}
( \beta_1(\xi) + i \gamma_1(\xi))  {\bf 1}_{N_1} &    &  0\\
   &  \ddots   &    \\
  0&  & ( \beta_{m}(\xi)  + i \gamma_m(\xi))  {\bf 1}_{N_m} 
\end{array}
\right)
\end{equation}
Since $\Phi$ commutes with itself and the other fields are vanishing, the equations of motion reduce to
\begin{align}\label{ftjanus}
\beta_i ''(\xi) + 2 \tanh \xi  \beta_i'(\xi)&=0 & \gamma_i ''(\xi) + 2 \tanh \xi  \gamma_i'(\xi) &=0
\end{align}
which is solved by
\begin{align}\label{eq:beta-gamma-i}
\beta_i (\xi)&=  b_i + c_i \tanh \xi &  \gamma_i (\xi)&= f_i + g_i \tanh \xi
\end{align}
This solution corresponds to an interface between two surface operators with different values of $\Phi$. Namely,
\begin{equation}\label{phifield-J-int}
\Phi_\pm = {e^{- i \psi} \over \sqrt{2}}
\left(
\begin{array}{ccc}
( b_1 + i f_1\pm (c_1+ig_1))  {\bf 1}_{N_1} &    &   0 \\
   &  \ddots &  \\
  0& & ( b_{m}  + i f_m\pm (c_m+ig_m))  {\bf 1}_{N_m} 
\end{array}
\right)
\end{equation}
Note that the linearized probe brane equation (\ref{probelin}) has the same form as the Yang-Mills equation (\ref{ftjanus}),
and it may be tempting to interpret the nonlinear corrections to the probe embedding in (\ref{ftjanus}) as strong coupling corrections to the semiclassical solution given above. 

The expectation value of the defect is again computed from the on-shell action.
The action (\ref{eq:N4-action}) reduces on shell to a boundary term, given by
\begin{align}
 S&=\frac{1}{g_{\rm YM}^2}\int d^4x\, \partial_\xi \left[\sqrt{g}\,{\rm tr}\left(\bar \Phi \partial_\xi \Phi\right)\right]
 =\frac{2\pi}{g_{\rm YM}^2}{\rm Vol}_{AdS_2}\left[\sqrt{g}\,{\rm tr}\left(\bar \Phi \partial_\xi \Phi\right)\right]_{\xi=-\infty}^{\xi=+\infty}
\end{align}
Using $\Phi$ in (\ref{phifield-J}) with $\beta_i$, $\gamma_i$ in (\ref{eq:beta-gamma-i}) 
now leads to a non-zero on-shell action
\begin{align}
 S&=\frac{2\pi}{g_{\rm YM}^2}{\rm Vol}_{AdS_2}\sum_{i}(c_i^2+g_i^2)N_i
\end{align}
Similar to the discussion below (\ref{eq:SD3-onshell}), finite counterterms could be added on the boundary of $AdS_3$, but are fixed to be absent by demanding the on-shell action to vanish for the supersymmetric configurations with constant $\beta_i$, $\gamma_i$.
The expectation value for the surface defect operator is thus given by
\begin{align}\label{eq:OSigma-N4}
 \langle \mathcal O_\Sigma\rangle&=\exp\left[-\frac{2\pi}{g_{\rm YM}^2}{\rm Vol}_{AdS_2}\sum_{i}(c_i^2+g_i^2)N_i\right]
\end{align}
Identifying the field theory and supergravity parameters via (\ref{beta-gamma}) leads to $\lambda\epsilon^2=(c_i^2+g_i^2)/(2\pi)^2$. 
The (renormalized) volumes of $AdS_2$ were given in (\ref{eq:VolAdS2}).
For the leading non-trivial order in $\epsilon$ and an interface separating two copies of global $AdS_2$, we thus find a factor $2$ discrepancy between the holographic computation at strong coupling, leading to (\ref{eq:cO-J}), and the semi-classical field theory computation leading to (\ref{eq:OSigma-N4}). 
We note in that context that even small $\epsilon$ amounts to large values for the scalar field in the field theory, due to the factor of $\sqrt{\lambda}$ in the identification (\ref{beta-gamma}), such that a semi-classical analysis for non-supersymmetric configurations may not be expected to be accurate at strong coupling.

\subsection{More general Janus on the D3-brane}\label{sec:gen-D3-Janus}

The ansatz of section~\ref{sec:janus} may be generalized by allowing the phase $\phi_0$ and the gauge field holonomy $A_\psi$ to dependent on the $AdS_2$ slicing coordinate $\xi$ as well
\begin{align}
u&=u(\xi) & \psi &= - \phi+ f(\xi)  & A_{\psi}&= a(\xi)
\end{align}
For this embedding the D3-brane action is given by
\begin{align}\label{genemb}
S_{\rm D3} &= T_{\rm D3}{\rm Vol}_{AdS_2}\int d\xi d\psi\, L_{\rm D3}
\nonumber\\
L_{\rm D3} &=\cosh^2\xi\, \left[\cosh^2\!u\sqrt{\cosh^4\!u + (a')^2+ \sinh^2\!u\;  (f')^2 + \cosh^2\!u \; (u')^2}-\cosh^4\!u+1\right]
\end{align}
The action depends on $f$ and $a$ only through their derivatives, such that $f'$ and $a'$ are determined in terms of $u$ by conservation laws.
Together with the equation of motion for $u$ following from the variation of (\ref{genemb}) 
this leads to Janus-type solutions, which interpolate between different constant values for $u$, $\phi_0$ and $A_\psi$ as $\xi\rightarrow\pm\infty$.
With the identification of these parameters with those of the surface operator given in section \ref{sec2}, these solutions realize an interface which interpolates between different values of $\beta,\gamma$ and $\alpha$.\footnote{The remaining parameter $\eta$ is associated with the holonomy of the dual gauge field, which we do not consider here.}

A perturbative solution can once again be obtained straightforwardly.
For solutions with
\begin{align}
 \lim_{\xi\rightarrow\pm\infty}\sinh u(\xi)&=\sinh u_0\pm\epsilon
 &
 \lim_{\xi\rightarrow\pm\infty}f(\xi)&=\phi_0\pm \delta f
 &
 \lim_{\xi\rightarrow\pm\infty}a(\xi)&=a_0\pm\delta a
\end{align}
where $\delta f$ and $\delta a$ are of $\mathcal O(\epsilon)$,
the first terms in the perturbative solution are 
\begin{align}
\sinh u=\,&\sinh u_0+\epsilon  \tanh \xi-\frac{\delta f^2 \sinh u_0}{2\cosh^2\!\xi}  +\mathcal O(\epsilon^3)
\nonumber\\
a=\,& a_0+\delta a \tanh \xi +\mathcal O(\epsilon^3)
\nonumber\\
f=\,&\phi_0+\delta f  \tanh \xi 
+\frac{\epsilon\,\delta f}{\cosh^2\!\xi\sinh u_0}
+\mathcal O(\epsilon^3)
\end{align}
These perturbative solutions are clearly of Janus form at the leading order, and this behavior again extends to the non-linear solutions.
The on-shell action evaluates to
\begin{align}
 S_{\rm D3, on-shell}= 
 S_{\rm D3}^{(0)}+2 \pi T_{\rm D3}  {\rm Vol}_{AdS_2}\Bigg\{&
 \epsilon^2+\delta a^2+\delta f^2 \sinh ^2\!u_0
 -\frac{2(\epsilon^2+\delta a^2+\delta f^2\sinh^2\!u_0)^2}{15\cosh^4\!u_0}
 \nonumber \\ &
 -\epsilon^2\delta f^2-\frac{1}{3}\delta f^4\sinh^2\!u_0
 +\mathcal O(\epsilon^6)
  \Bigg\}
\end{align}
For $\delta f=\delta a=0$ it reduces to (\ref{eq:cO-J}).
For $u_0=0$ the $S^1_\psi$ degenerates, which is reflected in the appearance of $\delta f$ in combination with $\sinh^2\!u_0$.
Within this more general ansatz for D3-brane embeddings it might be possible to find solutions which preserve some supersymmetry. In a preliminary analysis we found configurations that solve the non-linear equations of motion and are supersymmetric, but they are complex and their physical interpretation is unclear. These complex solutions are discussed briefly in appendix~\ref{sec:susy-D3}.

\section{Janus on other branes}\label{sec3}
The Janus-on-the-brane solution found in the previous section can be generalized to probe branes of other dimensions 
in different $AdS$ backgrounds.
In this section we discuss three cases: 
M2-branes in the $AdS_4 \times S^7$ solution of M-theory with ABJM as dual field theory, 
M5-branes in the $AdS_7\times S^4$ solution of M-theory with 6d $\cN=(2,0)$ theories as holographic duals, 
and D4-branes in the $AdS_6\times S^4/\ZZ_2$ vacuum of massive Type IIA found by Brandhuber and Oz \cite{Brandhuber:1999np} with 5d $USp(N)$ theories as dual. 

We will show that the aforementioned probe branes with $p$-dimensional worldvolume admit Janus-on-the-brane embeddings into the $AdS_{p+2}\times \mathcal M^{D-p-2}$ backgrounds. 
For the $AdS_{p+2}$ part of the background we will use an $AdS_{p}\times S^1$ slicing, with $AdS_p$ in turn sliced by $AdS_{p-1}$, such that the metric takes the form
\begin{align}\label{eq:ds2-AdS-p}
 ds^2_{AdS_{p+2}}&=du^2+\cosh^2\!u\,ds^2_{AdS_{p}}+\sinh^2\!u\, d\psi^2
 \nonumber\\
 ds^2_{AdS_p}&=d\xi^2+\cosh^2\!\xi\,ds^2_{AdS_{p-1}}
\end{align}
The probe branes can be embedded in such a way that they wrap $AdS_{p-1}\times S^1_\psi$ in $AdS_{p+1}$, and that upon going around the $S_\psi^1$ in $AdS_{p+1}$ they wind around an $S^1$ in the internal space. The remaining worldvolume coordinate is $\xi$, and the embeddings are characterized by a function $u(\xi)$.
We show that for appropriate winding numbers the brane Lagrangian reduces to 
\begin{equation}\label{eq:L-eff}
L= \tilde T_p \cosh^{p-1}\! \xi\, \Big[ \cosh^{p+1}\! u\left(\sqrt{1 + u'^2\sech^2\!u}-1\right)+1\Big]
\end{equation}
with a constant $\tilde T_p$, and with all other equations of motion satisfied.
The D3-brane in $AdS_5\times S^5$ discussed in the last section corresponds to $p=3$,
the M2-brane to be discussed in section~\ref{sec:M2} to $p=2$, 
the M5-brane to be discussed in section~\ref{sec:M5} to $p=5$, and the D4-brane to be discussed in section~\ref{sec:D4} to $p=4$.
Janus-on-the-brane solutions will be discussed based on this general form of the Lagrangian in section~\ref{sec:gen-Janus}.

\subsection{M2-brane in \texorpdfstring{$AdS_4\times S^7$}{AdS4xS7}}\label{sec:M2}

For a probe M2-brane in the $AdS_4\times S^7$ solution of M-theory we utilize  an $AdS_2\times S^1 $ slicing of $AdS_4$ and an $S_1\times S^5$ slicing of $S^7$,
\begin{equation}
ds^2=   L^2ds^2_{AdS_4} + 4L^2 \Big( d\theta^2 + \sin^2\theta d\phi^2 + \cos^2\theta ds_{S_5}^2 \Big)
\end{equation}
The metric on $AdS_4$ is given by (\ref{eq:ds2-AdS-p}) with $p=2$; the $AdS_{p-1}$ degenerates for this case and we simply have
\begin{align}
 ds^2_{AdS_2}&=d\xi^2-\cosh^2\!\xi\,dt^2
\end{align}
The three-form potential $C_3$ is given by
\begin{equation}
C_3=L^3 (\cosh^3\! u-1)  \vol_{AdS_2} \wedge d\psi
\end{equation}
The action for a single probe M2-brane  is given by
\begin{equation}\label{m2act}
S_{\rm M2}= T_2 \int d^3 \xi \sqrt{-\det( g)} - T_2 \int C_3
\end{equation}
The  world-volume coordinates of the M2-brane in static gauge are  $\xi,t, \phi$ and we choose the following ansatz for the  embedding
\begin{align}
 u&=u(\xi) & \phi&=\phi(\psi) & \theta&={\pi\over 2}
\end{align}
The action (\ref{m2act})  becomes
\begin{align}
S_{\rm M2}=L^3T_2 \int d\xi dt d\psi \,\cosh \xi\,\Big\{ & \cosh u(\xi) \sqrt{\cosh^2\! u(\xi) + u'(\xi)^2}\sqrt{\sinh^2\!u(\xi) + 4 \dot\phi(\psi)^2}
\nonumber\\
&- \cosh^3\! u(\xi)+1\Big\}
\end{align}
The equation of motion of $\phi$ is solved by 
\begin{equation}
\phi = {1\over 2}\psi +\phi_0
\end{equation}
This solution is the analog  of (\ref{bpswind}) for the D3-brane. Here the embedding of the brane into $S^1_{\phi}\times S^1_{\psi}$  winds twice around $S^1_{\phi}$. For other choices of windings no BPS solution  with constant $u$  exists.
The action for $u$  reduces to 
\begin{align}
S_{\rm M2}&=L^3T_2 \int d\xi dt d\psi \cosh \xi  \,\left[ \cosh^3\! u(\xi) \left(\sqrt{1 + u'(\xi)^2\sech^2\! u(\xi)} - 1\right)+1\right]
\end{align}
As advertised, this is of the form (\ref{eq:L-eff}) with $p=2$.
Solutions to the resulting equation of motion with constant $u(\xi)$, corresponding to a probe M2-brane with $AdS_2\times S^1$ worldvolume, have been identified in  \cite{Drukker:2008jm} with duals of vortex loop operators, mainly in the case of $S^7/Z_k$ orbifolds dual to ABJM theories. Janus-on-the-brane solutions will be discussed in section~\ref{sec:gen-Janus}.

\subsection{M5-brane in \texorpdfstring{$AdS_7\times S^4$}{AdS7xS4}}\label{sec:M5}

One can obtain an analogous construction for a probe M5-brane in $AdS_7\times S^4$, utilizing an $AdS_5\times S^1 $ slicing of $AdS_7$ and an $S_1\times S^2$ slicing of $S^4$,
\begin{eqnarray}
ds^2&=&  4  L^2 ds^2_{AdS_7} + L^2 \Big( d\theta^2 + \sin^2\!\theta\, d\phi^2 + \cos^2\!\theta\, ds^2_{S^2} \Big) \nonumber\\
F_4&=& 3 L^4    \sin\theta \cos^2\!\theta  \; d\theta\wedge d\phi\wedge \vol_{S^2}
\end{eqnarray}
with the $AdS_7$ metric given by (\ref{eq:ds2-AdS-p}) with $p=5$ and $\vol_{S^2}$ the volume form on unit-radius $S^2$.
The M5-brane action involves a WZ-coupling to the potential $C_6$  for the dual field strength $F_7=dC_6= *_{11} F_4$.
The potential is given by
\begin{equation}
C_6 = 2^6 L^6 (\cosh^6\! u-1) \vol_{AdS_5} \wedge \,d\psi
\end{equation}
where $\vol_{AdS_4}$ is the volume form of unit-radius $AdS_4$. 
The action for an M5-brane is given by \cite{Bandos:1997ui,Aganagic:1997zq}
\begin{eqnarray}
S_{\rm M5}&=& \int d^6 \zeta \Big( \sqrt{-\det(g_{mn}+ i \tilde H_{mn})} -  {\sqrt{-g}\over 4 \partial_m a \partial^m a}    \partial_l H^{*lmn} H_{mnp}\partial^p a\Big)
\nonumber\\
&& - \int \Big(  \hat C_6 +{1\over 2} F\wedge C^{(3)}\Big)
\end{eqnarray}
where $\hat C^6$ is the pull back of $C_6$ to the worldvolume of the M5-brane.
For an M5-brane in $AdS_7\times S^4$ with vanishing self-dual antisymmetric tensor field the action reduces to
\begin{equation}\label{ldbim5}
S_{\rm M5}=\int d^6 \zeta \Big( \sqrt{-\det(g)} - \hat C_6\Big)
\end{equation}
The M5 brane has worldvolume coordinates $\xi$, $\psi$ and those of $AdS_4$, and we choose the embedding
\begin{align}
u&=u(\xi) & \phi&= \phi(\psi) &  \theta&=\frac{\pi}{2}
\end{align}
The action  (\ref{ldbim5}) becomes
\begin{align} 
S_{\rm M5}=16 L^6 \Vol_{AdS_4} \int d\xi d\psi\, \cosh^4\!\xi\, \Big(& \cosh^4\!u(\xi) \sqrt{4\sinh^2\! u(\xi) +  \dot\phi(\psi)^2}\sqrt{\cosh^2\!u(\xi) + u'(\xi)^2} \nonumber \\
& -2\cosh^6\!u(\xi)+1 \Big) 
\end{align}
The equation of motion for $\phi$ is solved by
\begin{equation}
\phi =-2 \psi + \phi_0
\end{equation}
As in the previous cases, only this choice of winding leads to a BPS embedding with constant $u$.
The action for $u(\xi)$ becomes
\begin{equation} 
S_{\rm M5}=32\pi L^6 \Vol_{AdS_4} \int d\xi \cosh^4\!\xi\,  \left[\cosh^6\!u(\xi)\left(\sqrt{1 + u'(\xi)^2\sech^2\!u(\xi)}-1\right)+1\right]
\end{equation}
This is of the form (\ref{eq:L-eff}) with $p=5$, as advertised.
A solution with constant $u$ describes a codimension-two defect in the 6d $\cN=(2,0)$ theory, and Janus-on-the-brane solutions will be discussed in section~\ref{sec:gen-Janus}.

\subsection{D4-brane in \texorpdfstring{$AdS_6\times S^4/\ZZ_2$}{AdS6xHS4}}\label{sec:D4}
The Brandhuber-Oz background  \cite{Brandhuber:1999np} is a solution of massive Type IIA supergravity which has the form of $AdS_6$ warped over a half $S^4$.  
We will use it in the following form: with  $l_s=1$ the metric in string frame is given by
\begin{eqnarray}
ds^2&=&   \left({3 \over 2}\right)^{5\over 3} { \sqrt{q_4} \over  (C \; m \sin \alpha)^{1\over3} }\left\{  ds^2_{AdS_6} + {4\over 9} \Big( d\alpha^2+ \cos^2\!\alpha\, ds^2_{S^3}\Big)\right\}
\end{eqnarray}
and we take the $AdS_6$ metric as given by (\ref{eq:ds2-AdS-p}) with $p=4$.
The dilaton $\phi$ and five-form potential for the dual six-form field strength $F_6=dC^{(5)}$ are given by
\begin{align}
 e^{-\phi} &= {(q_4)^{1\over 4}\over C} \Big( {3\over 2} C m \sin\alpha\Big)^{5\over 6} 
 & 
 C^{(5)} &= {3^5 (q_4)^{3/2}  \over 2^5 C}(\cosh^5\!u-1)\,\vol_{AdS_4}\wedge \,d\psi
\end{align}
With the embedding
 \begin{align}
u&=u(\xi) & \phi &= {9\over 4} \psi+ \phi_0 & \alpha&=0 &  \theta_1&=\theta_2={\pi/2}
\end{align}
the Born-Infeld and WZ action
\begin{equation}
S_{\rm D4}= \int d^5 \zeta e^{-\phi} \sqrt{-\det{g}} - \int C^{(5)}
\end{equation}
produces the following action for the embedding function $u(\xi)$
\begin{equation}\label{eq:SD4}
S_{\rm D4}= {3^5 q_4^{3\over 2} \over 2^5 C } 2\pi\Vol_{AdS_3} \int d\xi \cosh^3\!\xi\,  \left[ \cosh^5\!u(\xi)\left(\sqrt{1 + u'(\xi)^2\sech^2\! u(\xi)} -1\right)+1\right]
\end{equation}
Note that  the probe brane is located at $\alpha=0$ where the dilaton blows up and the geometry is singular. However, as remarked in \cite{Penin:2019jlf} the D4-brane action is nevertheless well behaved. 
As advertised, the action (\ref{eq:SD4}) is of the form (\ref{eq:L-eff}) with $p=4$.
It would be interesting to investigate whether probe branes with an analogous form and corresponding solutions exist for the $AdS_6$ solutions of Type IIB supergravity constructed in \cite{DHoker:2016ujz,DHoker:2016ysh,DHoker:2017mds}.

\subsection{Janus on the brane for M2, D4, M5}\label{sec:gen-Janus}

We now discuss Janus-on-the-brane embeddings for the probe M2, D4 and M5 branes.
We have seen that the probe brane action reduces to (\ref{eq:L-eff}), which we repeat for convenience
\begin{equation}\label{eq:L-eff-2}
L= \tilde T_p \cosh^{p-1}\! \xi\, \left[\cosh^{p+1}\! u\, \left( \sqrt{1 + u'^2\sech^2\!u}-1\right)+1\right]
\end{equation}
The full equation of motion for $u$ reads
\begin{align}
\partial_\xi\left[\frac{u^\prime (\cosh\xi\cosh u)^{p-1}}{\sqrt{1+{u'}^2\sech^2\!u}}\right]
+\cosh^{p-1}\!\xi\,\cosh^p\!u\,\sinh u\left[p+1-\frac{1+p(1+{u'}^2\sech^2\!u)}{\sqrt{1+{u'}^2\sech^2\!u}}\right]&=0
\end{align}
It in particular admits (arbitrary) constant embeddings $u(\xi)=u_0$ as solutions.
Following the logic of section~\ref{sec:janus}, perturbative solutions can again be constructed using the ansatz 
\begin{equation}
\sinh u(\xi) = \sinh u_0 + \epsilon  u_1(\xi) + {1\over 2} \epsilon^2 u_2(\xi) +{1\over 3!} \epsilon^3 u_3(\xi) +\ldots 
\end{equation}
and the leading-order perturbation is determined by
\begin{align}
 u_1^{\prime\prime}(\xi)+(p-1)\tanh\xi\,u_1^\prime(\xi)&=0~.
\end{align}
This equation can be solved for general $p$ in terms of hypergeometric functions.\footnote{While the solutions for $p=2,3,4,5$ stand out in having a natural interpretation as probe brane embeddings, the equation can be studied for generic $p$. In fact, the solutions are of Janus form for generic $p>1$. For $p=1$ the solution is linear, for large $p$ it approaches a step function.}
The solution for $p=3$ was given in (\ref{lins}), and the solutions for the cases discussed in this section are
\begin{align}
 u_1&=\frac{4}{\pi} \alpha_1\tan ^{-1}\left(\tanh \frac{\xi }{2}\right)+\alpha_2 & p&=2
 \nonumber\\
 u_1&=\frac{4}{\pi} \alpha_1\tan ^{-1}\left(\tanh \frac{\xi }{2}\right)+\frac{2}{\pi}\alpha_1 \tanh \xi  \sech\xi +\alpha_2 & p&=4
 \nonumber\\
 u_1&=\frac{1}{2} \alpha_1\tanh \xi  \left(\sech^2\!\xi +2\right)+\alpha_2 & p&=5
\end{align}
These solutions are all of Janus form, interpolating between different finite values for $\xi\rightarrow\pm\infty$. 
For $\alpha_1=1$ and $\alpha_2=0$ they satisfy $\lim_{\xi\rightarrow \pm\infty}\sinh u(\xi)=\sinh u_0 \pm\epsilon$.
The Janus behavior extends to the non-linear solutions, in parallel to the discussion of section~\ref{sec:janus}.

Hence,  interfaces on codimension-two defects can be realized in a form similar to the D3-brane case.
The M2-brane in $AdS_4\times S^7$ describes a vortex operator in ABJM theory, and the Janus embedding corresponds to an interface point on this line defect.
The D4-brane describes a 3-dimensional defect in the 5d $USp(N)$ theories, and the Janus embedding corresponds to a 2-dimensional interface on the defect.
Lastly, the M5-brane describes a 4-dimensional defect in 6d $\cN=(2,0)$ theories, and the Janus embedding describes a 3-dimensional interface on the defect.
The computation of holographic observables can be done analogously to the D3-brane case.
A noteworthy feature is that the renormalized volume of $AdS_{p-1}$, appearing e.g.\ in the expectation value in (\ref{eq:cO-J}), is well defined only for odd-dimensional interfaces. For even-dimensional interfaces the scheme-independent information is in general in the logarithmic divergences, reflecting the presence of conformal anomalies. We leave more detailed studies for  future work.

\section{Discussion}\label{sec4}

We have presented non-supersymmetric deformations of $AdS_p$ probe brane embeddings that describe codimension-two defects in the dual field theory. The ansatz is based on an $AdS_{p-1}$ slicing of the $AdS_p$ part of the brane worldvolume, with the embedding dependent on the slicing coordinate. Remarkably, this ansatz works for half-BPS defects in all maximally supersymmetric $AdS_{p+2}\times S^q$ vacua of Type IIB and M-theory, as well as in the Brandhuber-Oz solution of massive Type IIA, and produces qualitatively similar solutions.   The equation determining the deformed solution is  a nonlinear ODE which can be solved numerically or perturbatively for small deformations of the supersymmetric embedding. In the field theory these branes  describe two halves  of $p-1$ dimensional defects, characterized by different values of the asymptotic embedding parameter, glued together at a $p-2$ dimensional interface.

A semi-classical field theory analysis as well as the $\kappa$-symmetry of the probe brane show that the Janus-on-the-brane solution breaks all supersymmetries. Since for large values of the slicing coordinate the solution approaches the supersymmetric embedding, we do not expect global instabilities. It would be interesting to investigate more systematically whether a supersymmetric generalization of the Janus-on-the-brane solution can be found.  For the original Janus solution in Type IIB supergravity \cite{Bak:2003jk} such solutions were indeed found in \cite{DHoker:2007zhm} and they are considerably more complicated than the nonsupersymmetric ones. The supersymmetric solutions reported in this paper are complex and their physical interpretation unclear, so one may have to consider more general embeddings.

Another interesting question  is wether it is possible to describe more complicated junctions of surface operators, which have been discussed for $\cN=4$ SYM  in a mathematical setting in \cite{Chun:2015gda} and from the localization perspective recently in \cite{Wang:2020seq},  using probe branes. Such brane configurations, if they exist, would be analogs of multi-Janus solutions in supergravity \cite{DHoker:2007hhe} which describe junctions of interfaces. 
Finally, it would be interesting to investigate whether there are fully backreacted solutions describing interfaces on defects, generalizing the fully backreacted Type IIB solutions for BPS surface defects constructed in \cite{Gomis:2007fi}. We leave these and other interesting questions for future work.

\section*{Acknowledgements}
M.G. is grateful to Daniel Roggenkamp for a conversation in the Summer of 2018 which originally motivated this work. 
The work of M.G.~was  supported, in part,  by the National Science Foundation under grant PHY-19-14412. 
The work of C.F.U.~was supported, in part, by the US Department of Energy under Grant No.~DE-SC0007859, by the Leinweber Center for Theoretical Physics, and by the Mani L. Bhaumik Institute for Theoretical Physics.

\appendix

\section{D3-brane supersymmetry}

We briefly discuss the supersymmetry of the  Janus-on-the-brane solution for the D3 brane in the $AdS_5\times S^5$ solution of Type IIB. We show that the Janus-on-the-brane solution constructed in section \ref{sec:janus} breaks all supersymmetries and briefly discuss complex supersymmetric solutions within the ansatz of section~\ref{sec:gen-D3-Janus}.
The supersymmetries preserved by the D3-brane are singled out by a constraint on the $AdS_5\times S^5$ Killing spinors $\epsilon$ \cite{Bergshoeff:1996tu, Cederwall:1996pv,Cederwall:1996ri},
\begin{align}
 \Gamma_\kappa\epsilon&=\epsilon
\end{align}
We use complex notation for the Killing spinors with conventions as in \cite{Karch:2015vra,Robinson:2017sup}.
For $AdS_5$ in $AdS_3\times S^1$ slicing and the $S^5$ in $S^3\times S^1$ slicing we use coordinates such that
\begin{align}
 ds^2_{AdS_5}&=du^2+\cosh^2\!u\,\left(dr^2+\cosh^2\!r\, ds^2_{AdS_2}\right)+\sinh^2\!u\,d\psi^2
 \nonumber\\
 ds^2_{S^5}&=d\theta^2+\sin^2\!\theta\,ds^2_{S^3}+\cos^2\!\theta d\phi^2
\end{align}
where $\theta$ has been shifted compared to (\ref{ads5met}), and
\begin{align}
 ds_{S^3}^2&=d\chi_1^2+\sin^2\!\chi_1 (d\chi_2^2+\sin^2\!\chi_2 d\chi_3^2)
 &
 ds^2_{AdS_2}&=dx^2-e^{2x}dt^2
\end{align}
The Killing spinors are given by
\begin{align}\label{eqn:Killing-R}
 \epsilon&=\RAdS\RS \epsilon_0
\end{align}
where, with $\GammaChi:=\Gamma^{\underline{\chi_1}}\Gamma^{\underline{\chi_2}}\Gamma^{\underline{\chi_3}}$, 
\begin{align}\label{rfivemat}
   R_{\mathrm{S}^5}&=
   e^{\frac{\theta}{2} i\Gamma_{}^{\underline{\phi}}\GammaChi}
   \:e^{\frac{\phi}{2} i\GammaChi\Gamma^{\underline{\theta}}}
   \:e^{\frac{1}{2}\chi_1\Gamma^{\underline{\theta\chi_1}}}
   \:e^{\frac{1}{2}\chi_2\Gamma^{\underline{\chi_1\chi_2}}}
   \:e^{\frac{1}{2}\chi_3\Gamma^{\underline{\chi_2\chi_3}}}
\\
  \RAdS&=e^{\frac{i}{2}u\Gamma_{\underline{u}}\GammaAdS}\:e^{\frac{1}{2}\psi\Gamma_{\underline{u\psi}}}\:e^{\frac{i r}{2}\Gamma_{\underline{r}}\GammaAdS} R_{\mathrm{AdS}_2}
  \nonumber\\
  R_{\mathrm{AdS}_2}&=e^{\frac{i x}{2}\Gamma_{\underline{x}}\GammaAdS}
 +\frac{i}{2}te^{\frac{x}{2}}\Gamma_{\underline{t}}\GammaAdS (\mathds{1}- i\Gamma_{\underline{x}}\GammaAdS)
\label{rads5}
\end{align}

\subsection{Janus on the brane embedding}\label{sec:job-susy}

In the simplest Janus-on-the-brane solution discussed in section~\ref{sec:janus} the D3-brane wraps $AdS_3$ with coordinates $(r,t,x)$ in $AdS_5$ and the $S^1$ with coordinate $\psi$ in $S^5$. We can redefine coordinates to set the shift in (\ref{bpswind}) to zero and take, without loss of generality, 
\begin{align}
 u&=u(r) & \psi&=\phi & \theta&=\chi_i=0
\end{align}
The pullback of the Clifford-algebra matrices to the D3-brane, $\gamma_i=E_\mu^a(\partial_i X^\mu)\Gamma_a$, is given by
\begin{align}\label{wvgamma}
 \gamma_r&=\cosh u \; \Gamma_{\underline{r}}+u^\prime\Gamma_{\underline{u}} &
   \gamma_\phi&=\Gamma_{\underline{\phi}}+\sinh u \; \Gamma_{\underline{\psi}}
   \nonumber\\
 \gamma_t&= \cosh u\cosh r e^x \; \Gamma_{\underline{t}} & \gamma_x&=\cosh u\cosh r  \; \Gamma_{\underline{x}}  
\end{align}
and the induced metric on the D3-brane is
\begin{align}
 g&=\left(\cosh^2\!u+u'^2\right)dr^2+\cosh^2\!u \left(d\phi^2+\cosh^2\!r\,ds^2_{AdS_2}\right)
\end{align}
The $\kappa$-symmetry constraint for this embedding is
\begin{align}
 \Gamma_\kappa\epsilon&=\epsilon &  \Gamma_{\kappa}&=\frac{1}{\sqrt{-\det(g)}}\gamma_{rtx\phi}
\end{align}
For $\theta=\chi_i=0$, the matrix $R_{S^5}$  defined in (\ref{rfivemat}) simplifies on the D3-brane worldvolume to
\begin{align}\label{eqn:S5-R-matrix}
   R_{\mathrm{S}^5}&=e^{\frac{\phi}{2} i\GammaChi\Gamma^{\underline{\theta}}}
\end{align}
and as a result the $\kappa$-symmetry condition simplifies to
\begin{align}\label{eq:kappa-1}
 -i\gamma_{r\phi}\Gamma_{\underline{tx}} \RAdS\epsilon_0&=h\RAdS \epsilon_0
 &
 h&=\cosh u \,\sqrt{\cosh^2\!u+(u^\prime)^2}
\end{align}

\subsubsection{\texorpdfstring{$\kappa$}{kappa}-symmetry}

To show that the Janus-on-the-brane embedding of section~\ref{sec:janus} is not supersymmetric we set $\psi=t=x=0$.
The $\kappa$-symmetry condition (\ref{eq:kappa-1}) becomes
\begin{align}
e^{-\frac{i}{2}u\Gamma_{\underline{u}}\GammaAdS}\:e^{-\frac{i r}{2}\Gamma_{\underline{r}}\GammaAdS}( \cosh u \Gamma_{\underline{r}} + u' \Gamma_{\underline{u}})(\Gamma_{\underline{\phi}}+\sinh u \Gamma_{\underline{\psi}})
\Gamma_{\underline{tx}}
e^{\frac{i}{2}u\Gamma_{\underline{u}}\GammaAdS}\:e^{\frac{i r}{2}\Gamma_{\underline{r}}\GammaAdS}  \epsilon_0&=i h \epsilon_0
\end{align}
Evaluating this expression more explicitly leads to
\begin{align}
 \cosh u\sinh u\left(\cosh r \left(i\Gamma_{\underline{\phi\psi}}+\Gamma_{\underline{rtx\psi}}\right)-\sinh r \left(i\Gamma_{\underline{ru}}+\Gamma_{\underline{utx\phi}}\right)\right)\epsilon_0
 &\nonumber\\
 +\left(\cosh^2\!u\,\Gamma_{\underline{r tx\phi}}-ih\mathds{1}\right)\epsilon_0
 &
 \nonumber\\
 +u'\left(\left(\cosh r \Gamma_{\underline{u}}+i \sinh r \Gamma_{\underline{tx\psi}}\right)
 \left(\Gamma_{\underline{tx\phi}}+i\sinh^2\!u\,\Gamma_{\underline{r}}\right)
 +\sinh u\cosh u \Gamma_{\underline{utx\psi}}\right)\epsilon_0&=0
\label{kappa-b}
\end{align}
Consider now the limit $r\to \pm  \infty$ where $\lim_{r\to \pm \infty} u = u_{\pm}$ and $\lim_{r\to \pm  \infty} u'  e^{|2r|}=const$. 
The leading terms in the first line of (\ref{kappa-b}) are $\mathcal O(e^{|r|})$, those of the second line are $\mathcal O(1)$ and those of the third line $\mathcal O(e^{-|r|})$.
At leading order, $\mathcal O(e^{|r|})$, the $\kappa$-symmetry condition becomes
\begin{align}
ie^{|r|}\cosh u_\pm \sinh u_\pm \Big( \Gamma_{\underline{\phi \psi}}\mp\Gamma_{\underline{ru}} \Big)\Big(1+i \Gamma_{\underline{r t x \phi}}\Big) \epsilon_0 &=0
\end{align}
This condition is satisfied if
\begin{align}\label{eq:susy-const}
 -i\Gamma_{\underline{rtx\phi}} \epsilon_0&= \epsilon_0
\end{align}
In fact, the entire first line in (\ref{kappa-b}) vanishes with this constraint.
Since $h\to \cosh^2\!u\,+\mathcal O(e^{-2|r|})$, the second line of (\ref{kappa-b}) reduces to $\mathcal O(e^{-2|r|})$.
For constant embeddings, the second and third lines of (\ref{kappa-b}) vanish altogether, showing that the constant embedding preserves the supersymmetries characterized by (\ref{eq:susy-const}) and is $\tfrac{1}{2}$-BPS.
For non-constant embeddings the next non-trivial order in (\ref{kappa-b}) is $\mathcal O(e^{-|r|})$, due to terms in the third line.
The condition at that order becomes
\begin{align}
e^{|r|} u^\prime \cosh^2 u_{\pm}  \Big( \mp \Gamma_{\underline{\phi\psi}} +\Gamma_{\underline{ru}}\Big) \epsilon_0 &=0
\end{align}
It implies that for non-zero $u^\prime$ there is no consistent projection condition which makes the terms of order $e^{-|r|}$ in the $\kappa$-symmetry condition vanish both for large positive and negative $r$. Hence the Janus-on-the-brane embedding of section~\ref{sec:janus} breaks all supersymmetries.

\subsection{Supersymmetric embeddings}\label{sec:susy-D3}

The embedding ansatz of section~\ref{sec:janus} can be generalized as in section~\ref{sec:gen-D3-Janus}, and within this generalized ansatz we indeed found supersymmetric embeddings. They are complex, however, making their physical interpretation unclear, and we will present them without derivation.
They are characterized by
\begin{align}
 u'(\xi)&=\pm \sech^2\!\xi\, \sech u(\xi ) \sqrt{-p^2 \csch^2\! u(\xi)-q^2}
 \nonumber\\
 f'(\xi)&=p\sech^2\!\xi\,\csch^2\!u(\xi)
 \nonumber\\
 a'(\xi)&=q \sech^2\!\xi
\end{align}
with constants $p$, $q$.
The first equation can be integrated for $u(\xi)$, and then $a(\xi)$, $f(\xi)$ are given explicitly.
These configurations solve the full non-linear equations of motion derived from (\ref{genemb}), and some of them are supersymmetric, for example for $(p,q)=(\pm 1,0)$.
However, one can not make the embedding function $u$, the relation between $\phi$ and $\psi$,  and the flux on the D3-brane all real at the same time. We leave a physical interpretation open and a more exhaustive analysis of $\kappa$-symmetry for more general embeddings for the future.

\bibliographystyle{JHEP}
\bibliography{job}

\end{document}